\documentclass[prb,showpacs,preprint]{revtex4}   
\usepackage{xspace}
\usepackage{times,mathptm}                            
\usepackage{graphicx}                            
\usepackage{epstopdf}
\usepackage{amsmath}
\usepackage{amssymb}
\newcommand{\gra}{$^{\circ}$\xspace}
\newcommand{\celsius}{$^{\circ}$C\xspace}

\newcommand{\etal}{\emph{et al.}\xspace}

\begin{document}

\title{Two dimensional confinement of electrons in nanowall network of GaN leading to high mobility and phase coherence}
\author{H.\ P.\ Bhasker$^{1}$}
\author{Varun\ Thakur$^{2}$, S.\ M.\ Shivaprasad$^{2}$}
\author{S.\ Dhar$^{1}$}
\email{dhar@phy.iitb.ac.in}
\affiliation{Physics Department, Indian Institute of Technology Bombay Powai, Mumbai 400076, India}
\affiliation{International Centre for Material Science, Jawaharlal Nehru Centre for Advanced Scientific Research, Bangalore 560064, India}

\begin{abstract}

Here, we report an alternative route to achieve two dimensional electron gas (2DEG) in a semiconductor structure. It has been shown that charge accumulation on the side facets can lead to the formation of 2DEG in a network of c-axis oriented wedge-shaped GaN nanowalls grown on c-plane sapphire substrate. Our study reveals that negative charges on the side-facets  pushes the electron cloud inward resulting in the formation of 2DEG in the central plane parallel to the wall height. This confinement is evidenced from several orders of magnitude enhancement of electron mobility as compared to bulk, observation of weak localization effect in low temperature magneto-transport studies as well as the reduction of both the elastic and inelastic scattering rates with the average width of the walls. Importantly, the phase coherence length has been found to be as high as 20~$\mu$m, which makes the system potentially interesting for spintronics. Schr$\ddot{\mathrm{o}}$dinger and the Poisson equations are solved self-consistently taking into account the surface charge accumulation effect. The result indeed shows the 2D quantum confinement of electrons even for 40~nm of wall-width.

\end{abstract}

\maketitle
Two dimensional (2D) carrier gas, where the carriers are virtually confined in a 2D sheet, offers a test bed to explore new directions in physics by giving rise to many remarkable phenomena such as several orders of magnitude enhancement of mobility, integer and fractional quantum hall effects\cite{klitzing,tsui,laughlin},  2D metal insulator transition\cite{abrahams} and microwave induced zero resistance state\cite{mani}. The system thus remains to be the subject of intense research for several decades. Most of the techniques to achieve 2D confinement of carriers involve semiconductors. The common approach is to confine carriers either in a triangular potential well formed at the heterointerface between two semiconductors or in a rectangular potential well formed by sandwiching a lower band gap semiconductor layer between higher band gap semiconductors.\cite{mitin} These heterostructure based 2D systems made of only a few limited semiconducting materials, mainly GaAs, are extensively investigated so far. Many of the fascinating results mentioned above are in fact observed in these systems. Recently, new classes of 2D systems have opened up novel aspects of 2D carriers. For example, the formation of high electron mobility 2D electron gas at the interface between two insulating oxides - LaAlO$_3$ and SrTiO$_3$.\cite{ohtomo} The 2DEG, in this case, arises as a result of electron entrapment in certain exotic interface states, which are not possible to realize in bulk.  Another example is graphene, where 2D carrier gas is formed as a result of atomically thin layer width.\cite{novoselov,zhang} In graphene, electrons behave like massless charge particles due to linear dispersion in the band profile. 

Here, we report an entirely different route to achieve 2D quantum confinement of electrons wedge-shaped GaN nanowalls. Recently, electron mobility in networks of c-axis oriented GaN nanowalls is estimated to be several orders of magnitude larger than that is observed in GaN bulk.\cite{bhasker1} The origin of this effect was not very clear. It was speculated to be resulting from the transport of electrons through the edge states formed at the top edges of the nanowalls.\cite{bhasker1} Here, we have studied the surface fermi-level position using x-ray photoelectron spectroscopy (XPS) as well as depth distribution of transport, and magneto-transport properties of the nanowalls to understand the real cause of the effect. The study suggests that the accumulation of negative charges on side-facets of the walls pushes the electrons inward, leading to the formation of 2DEG in the central plane parallel to the height of the wall. This confinement leads to several orders of magnitude reduction of both elastic and inelastic scattering rates as compared to bulk, resulting in a significant enhancement in electron mobility and weak localization effect. Our study furthermore reveals that both the elastic and inelastic scattering rates decreases with the average width of the walls, providing a strong evidence for the 2DEG formation. In order to theoretically validate such a scenario, we have taken into account the surface band bending due to the accumulation of negative charges on both the surfaces of a infinitely extended GaN slab and solved the Schr$\ddot{\mathrm{o}}$dinger and Poisson (SP) equations self-consistently to obtain the potential and charge density distribution across the wall-width. The result indeed supports the 2D quantum confinement of electrons even for 40~nm of wall-width.

It should be mentioned that we have investigated several nanowall network samples and obtained qualitatively similar results in all of them. However, for the sake of clarity, we will concentrate here on a representative sample, which was grown directly on a c-plane sapphire substrates using plasma assisted molecular beam epitaxial (PA-MBE). The growth was carried out for four hours at a substrate temperature of 640\celsius, Ga flux of 3.86 $\times$ 10$^{14}$ cm$^{-2}$s$^{-1}$ and nitrogen flow rate of 4.5 sccm. Details of the growth and structural characterization of these layers are reported elsewhere.\cite{growth}. Morphology of the sample was investigated using field emission scanning electron microscopy (FESEM). XPS was carried out to determine the band bending properties at the side facets of the nanowalls. In order to maximize the information coming from the side facets of the walls, the surface normal of the sample was kept at an angle of 70~\gra with respect to the incident electron beam. Magnetoresistance measurements were carried out at 2K in a liquid He cryostat equipped with a superconducting magnet. Conductivity measurements were carried out with four probe van-der-pauw contact geometry using Indium contact pads, which provided good ohmic contacts. 4M aqueous solution of KOH is used to etch the nanowalls for different time durations to study the depth distribution of their transport and magneto-transport properties.


 Figure \ref{fig1}(a) and (b) show the top view FESEM images of the nanowalls before and after 60 minutes of dipping in 4M aqueous solution of KOH, respectively. Nanowalls form well connected network structure, which remains intact even after the KOH treatment. However, the average tip width of the walls, which is estimated to be $\approx$ 15~nm prior to the dipping\cite{expl}, increases to $\approx$ 50~nm after dipping, evidencing the etching of the wedge-shaped nanowalls from the top. The insets of the panels (a) and (b) present the corresponding cross sectional FESEM images, which demonstrate the reduction of wall height with etching time. In Fig. \ref{fig1} (c), the height of the walls $h$ is plotted as a function of etching time, which yields an etching rate of $\approx$ 10~nm/min. The average tip-width $t_{av}$ and the area fraction $\phi$ occupied by the tip of the walls, which are obtained by analyzing several top view SEM images taken from different parts of the sample at various etching steps, are plotted as a function of the etching time in Fig. \ref{fig1} (d). Both $\phi$ and $t_{av}$ increase with the etching time, which is consistent with the tapered structure of the nanowalls. 


Figure \ref{fig2}(a) shows the four-probe resistivity $\rho_{FP}$ measured at 300~K as a function of the height of the nanowalls $h$. Interestingly, $\rho_{FP}$ increases by an order of magnitude as the wall height decreases from 1.2~$\mu$m to 0.2~$\mu m$. Since the average tip-width $t_{av}$ has been found to increase with the decrease of $h$ [see fig. \ref{fig1}(d)], the above observation is consistent with our earlier work\cite{bhasker2}, where the samples grown with different values of $t_{av}$ are found to show a reduction of conductivity with increasing tip width. In the upper inset of Fig.\ref{fig2}(a) the logarithm of  conductance $G$ = $h$/$\rho_{FP}$ is plotted as a function of $h$. Clearly, $\ln(G)$ versus $h$ data show a linear behavior, suggesting an exponential increase of $G$ with $h$. Note that $G$ would be the conductance between two parallel contact pads placed with a gap equal to their width for a uniform layer (100$\%$ coverage).  Since, in these nanowall networks, surface is not fully covered, $\rho_{FP}$ is not the actual resistivity of the walls. In fact, the wall resistivity is expected to be less than $\rho_{FP}$. Let us consider two parallel contact pads on the sample surface as shown schematically in the lower inset of Fig.~\ref{fig2}(b). The effective width of the contact pad, which will be in contact with the tip of the walls $l_{eff}$ = $\phi l/\pi$.\cite{bhasker1} Now, $G(h)$ = $\int_{0}^{h}\sigma(z)l_{eff}dz/l$, where $\sigma(z)$ is the conductivity at a height $z$ of the wall. $\sigma(h)$ = $\frac{\pi}{\phi(h)}\frac{dG}{dh}$, where $\frac{dG}{dh}$ can be obtained by fitting the $\ln(G)$ versus $h$ data with a straight line as shown in the inset of Fig. \ref{fig2}(a). In Fig.~\ref{fig2}(b), $\sigma$ is plotted as a function of $t_{av}$. Evidently, the top part of the walls is significantly more conductive than the bottom part. More notably, the conductivity at the tip of the walls is substantially more than what is expected for bulk GaN at room temperature. 


In order to verify whether the decrease of conductivity with etching is associated with the reduction of mobility, a field effect transistor device is fabricated as schematically shown in Fig.~\ref{fig3}(a) to independently measure mobility in nanowalls as a function of etching duration. A 80~$\mu$m mica sheet, which acts as the dielectric layer, is sandwiched between the GaN nanowall layer and a gold coated n-type Si layer by applying pressure. Fig.~\ref{fig3}(b) and (c) shows the source to drain current $I_{SD}$ as a function of gate to drain voltage $V_{GD}$ measured at 300~K with a fixed source-to-drain voltage $V_{SD}$ for the sample before and after 110~min of etching, respectively.  In both the cases, $I_{SD}$ increases as $V_{GD}$ increases beyond a threshold voltage $V_{th}$ in the positive direction, which implies that electrons are the majority charge carriers even after etching. The electron mobility $\mu$ can be expressed as  $\mu$ = $L^2 g_m$/$C_{GD} V_{SD}$\cite{fet}, where $g_m$ = $\frac{dI_{SD}}{dV_{GD}}$ is the transconductance and $C_{GD}$ is the gate-drain capacitance. $\mu$ can thus be estimated as 1.1$\times$10$^4$ and 2.1$\times$10$^2$cm$^2$/V-sec for the unetched and etched samples, respectively. Note that the electron mobility in unetched walls is several orders of magnitude larger than the theoretical mobility limit for bulk GaN, which is consistent with our previous results.\cite{bhasker1,bhasker2} 


Figure~\ref{fig4}(a) compares the change in conductance $\Delta G(B)$ = $G(B)-G(0)$ measured at 2~K as a function of the magnetic field $B$ for the unetched and the 60~min etched sample. Figure~\ref{fig4}(b) shows $\Delta G(B)$ plot in close-up for the 60~min etched sample. In both the cases, an increase of conductance with increasing magnetic field $\left|B\right|$ (negative magneto-resistance) is clearly visible at low fields. This can be attributed to the weak localization effect (WL), where a negative correction to the conductance arises due to the enhancement of the effective scattering cross-section as a result of constructive quantum interference between a closed path of electron and its time reversed path. When a magnetic field is applied, it reduces the negative correction by removing the phase coherence between the two paths.\cite{bergmann} It is noticeable that around $B$ = 0, the slope of $\Delta G(B)$ for the unetched sample is more than that for the etched one. In fact, the slope of $\Delta G(B)$ is found to decrease as the duration of etching increases. Observation of weak localization effect indicates a two dimensional nature of the transport.\cite{expl1,bergmann} It should be mentioned that the weak antilocalization has been reported in Group III-nitride heterostructure based 2DEG systems\cite{thillosen,schmult,lehnen}, where it is attributed to spin-orbit coupling arising through Rashba mechanism.\cite{schmult} These results thus suggest that the  spin-orbit coupling is insignificant in our sample.  Note that neither weak localization nor weak antilocalization has so far been observed in bulk phase of these materials. 
 

If electrons are quantum mechanically confined in a 2D channel, its mobility could be enhanced by several orders of magnitude due to the decrease in scattering cross-section as a result of reduced dimensionality, which could explain above results. However, it is difficult to conceive the idea of quantum confinement in these walls, which have an average width more than 20~nm and when the excitonic Bohr radius of GaN is $\approx$ 2.7~nm.\cite{hanada} Fig.~\ref{fig5}(a) compares the Ga 3$d$ XPS spectra for the nanowall sample and a GaN epitaxial layer grown on c-plane sapphire substrate in the same growth chamber. Interestingly, the peak position, which represents the distance of the Ga 3$d$ level from the surface fermi-level $E_{fs}$, is at 21.3~eV for the epilayer sample while it is only at 17.9 for the nanowall sample. Note that the observed position for Ga 3$d$ level for the epitaxial GaN sample matches well with reported values for n-type GaN layers.\cite{tracy} Since the position of the Ga 3$d$ level from the valence band is measured to be 17.76~eV\cite{waldrop}, the above finding indicates that while, the surface fermi-level $E_{fs}$ is located $\approx$ 100~meV above the conduction band minimum for the epilayer sample, it is lying $\approx$ 3.3~eV below the conduction band for the nanowall sample. This suggest the formation of positive depletion regions at the facets of the nanowall as shown schematically in Fig.~\ref{fig5}(b). Note that XPS study shows similar result in all of our nanowall samples. These observations lead us to believe that negative charges must be accumulated on the nanowall facets, which might be pushing the electrons inward leading to positive depletion regions at the boundaries and a 2D quantum confinement in the central plane parallel to the height of the wall. In order to examine such a plausibility, we have solved the Schr$\ddot{\mathrm{o}}$dinger and Poisson equations self-consistently satisfying the total charge neutrality within the wall. In the calculation, the position of the conduction band minimum at the two side faces of the wall is considered to be at 3.3~eV above the fermi level as observed in XPS study. Since the thermoelectric power measurement shows an electron concentration of 1 $\times$ 10$^{19}$ cm$^{-3}$ in this sample\cite{bhasker1}, we have taken donor concentration to be 1 $\times$ 10$^{19}$ cm$^{-3}$ in the calculation.  Donor activation energy is considered to be 30~meV as it is well known that oxygen acts as unintentional shallow donor with the same activation energy in GaN.\cite{ptak} Fig.~\ref{fig5}(c) compares the calculated conduction band profile inside a wall of thickness $w$ = 20 and 40~nm. Here we have considered the effective width $w$ of the nanowalls to be 20 and 40~nm before and after 60~min of etching , respectively. The charge density profiles are plotted in the inset of the figure, which indeed shows the formation of a 2DEG channel of thickness (negatively charged region at the center) $\approx$ 6~nm and 13~nm for $w$ = 20~nm and 40~nm, respectively. The energy eigenvalues $E_i$ and 2D electron density $n_{2D}$ for the two values of $w$ are also provided in Tab.~1.

Since the magnetic field is applied perpendicular to the sample surface in magnetoresistance measurements, the walls are always under a horizontal magnetic field. Beenakker and Houten have shown that for such a situation, the quantum correction to the conductance as a function of the magnetic field can be expressed as\cite{beenakker} 
\begin{equation}
\Delta G  = -N_{ch} \frac{e^2}{2 \pi^2 \hbar} \ln \left [ \left (\frac{\tau_e}{\tau_{\phi}}+\frac{\tau_e}{\tau_{B}} \right )^{-1} + 1 \right ]
\label{eq1}
\end{equation}      

Here, we have added the factor $N_{ch}$, which is the effective number of parallel 2D channels connecting the two contact pads.  $\tau_e$ is the mean free time between two elastic collisions, $\tau_{\phi}$ the phase coherence time and $\tau_{B}$ the magnetic field dependent phase coherence time. $\tau_{B}$ = $C_1 \hbar^2$/$w^3 e^2 B^2 v_f$ in weak magnetic field regime [$l_m$ $\gg$ $\sqrt{w l_e}$], where $C_1$ = 16 and 12.1 for diffuse and specular surface scattering, respectively, $v_f$ the fermi velocity, $l_e$ = $\tau_e v_f$ the mean free path and $l_m$ = $\sqrt{\hbar/e B}$ the magnetic length. $v_f$ can be obtained from the position of the fermi level with respect to the ground state calculated theoretically (see Tab.~1). $\tau_e$ has been obtained from the mobility measured at 300~K [see Fig.~\ref{fig3}]. Note that conductivity does not show any significant variation with temperature down to 2~K in this sample. We have thus considered the mobility at 2~K to be same as that at 300~K. These values  are listed in Tab.~1. It should be noted that the above expression for $\tau_{B}$ is valid as long as $l_e$ $\gg$ $w$. $l_e$ is estimated to be $\approx$ 720 and 19~nm for unetched and 110~min etched samples, respectively, justifying to use of the above mentioned expression for $\tau_{B}$ for the unetched sample. However, the etched sample belongs to the dirty regime [$l_e$ $\ll$ $w$] with $\tau_{B}$ = $4 \hbar^2$/$w^2 e^2 B^2 D$. $D$ is the diffusion coefficient, for which one has Fuchs formula for a thin film with diffuse boundary scattering $D$ = $\frac{1}{3} v_f l_e \left [ 1 - \frac{3l_e}{2W} \int_o^1{s(1-s^2)(1-exp(-w/sl_e) ds} \right]$.\cite{fuchs} Finally, the experimental data for $\Delta G(B)$, are fitted using Eqn.~\ref{eq1} with only $N_{ch}$ and $\tau_{\phi}$ as fitting parameters. Here, the surface scattering of the electrons, is considered to be diffuse type. The results are shown in Fig.~\ref{fig4}. The best fit values for $N_{ch}$ and $\tau_{\phi}$ are also provided in Tab.~1. The phase coherence length $l_{\phi}$ has been estimated from $\tau_{\phi}$ [$l_{\phi}$ = $\sqrt{D \tau_{\phi}}$].\cite{beenakker}  Evidently, the inelastic scattering rate for electrons, which governs $\tau_{\phi}$, increases with $t_{av}$, which is consistent with the formation of 2DEG. Interestingly, $\tau_{\phi}$ and $l_{\phi}$ for the unetched sample are higher than those reported for GaN/AlGaN heterostructure 2DEG\cite{thillosen,schmult,lehnen}, which suggests that the inelastic scattering rate is significantly lower in our sample than heterostructure 2DEG systems. This finding along with the absence of significant spin-orbit coupling makes this material potentially interesting for spintronics. 

It should be mentioned that our sample does not show Shubnikov de Haas (SdH) oscillation in the magnetoresistance down to 1.8~K for the magnetic field up to 9~T applied perpendicular to the substrate, meaning parallel to the walls. Note that SdH oscillation is not expected for the magnetic field oriented parallel to the 2DEG plane. It should also be mentioned that if the field is applied parallel to the substrate surface, it can never be perpendicular to all the walls at the same time due to the randomness in the in-plane arrangement of the walls.         

It has been found that several orders of magnitude enhancement of electron mobility as compared to bulk in networks of c-axis oriented wedge-shaped GaN nanowalls is resulting from 2DEG formed in the walls. Our study suggests that negative charges accumulate on the side-facets of the walls, which pushes the electron cloud inward leading to the 2D quantum confinement of electrons  in the central plane parallel to the height of the wall. Notably, the spin-orbit coupling as well as inelastic scattering rate in nanowall sample are found to be significantly lower as compared to heterostructure 2DEG system, which makes it potentially interesting for spintronics. Schr$\ddot{\mathrm{o}}$dinger and Poisson equations are solved self-consistently taking into account the surface charge accumulation effect. The result indeed supports the 2D confinement of electrons even for 40~nm of wall-width.

We acknowledge the financial support of this work by the Department of Science and Technology of the Government of India. SMS would like to thank Prof. C.\ N.\ R.\ Rao for his support and guidance.

\newpage
\phantom{000}
\begin{table}[t!]
\centering
\caption{Summary of the parameters for the GaN nanowall sample before and after 60~min of etching in aqueous solution of KOH: The wall width $w$, mobility $\mu$ obtained in Fig.~\ref{fig3}(b), energy eigenvalues $E_i$ and 2D electron concentration $n_{2D}$ obtained from self-consistent SP solution, Fermi velocity $v_f$ estimated from $E_i$, mean time between two elastic collisions $\tau_e$ [estimated from $\mu$], the mean free path $l_e$[=$\tau_{e} v_f$]. The phase coherence time $\tau_{\phi}$, and effective number of parallel 2D channels $N_{ch}$ are extracted from the fit using Eqn.~\ref{eq1}. The phase coherence length $l_{\phi}$ is obtained from $\tau_{\phi}$.}
\begin{tabular}{c|c|c|c|c|c|c|c|c|c|c} 
\hline
\hline

Sample & $w$  & $\mu$    & $E_{i}$   & $n_{2D}$  & $v_{f}$ & $\tau_{e}$ & $l_{e}$ & $N_{ch}$  & $\tau_{\phi}$ & $l_{\phi}$ \\ 
       &(nm) & (cm$^2$/V s) & (eV) & (cm$^{-2}$) & (M/s)  & (s)   & (nm)  & $cm^{-1}$  & (s)          & ($\mu$m)   \\ \hline
unetched  & 20   & 1.1 $\times$ 10$^4$ & -0.1892 & 1.6 $\times$ 10$^{13}$ & 5.8 $\times$ 10$^5$ & 1.25$\times$ 10$^{-12}$ & 723   &  3000        & 3.6$\times$ 10$^{-8}$ & 20.5 \\    
etched    & 40   & 2.1 $\times$ 10$^2$ & -0.2737 & 3 $\times$ 10$^{13}$   & 8 $\times$ 10$^5$   & 2.3$\times$ 10$^{-14}$  & 19    &  3000        & 8$\times$ 10$^{-11}$  & 0.57 \\ 
    &    &  & -0.0896 &    &    &   &     &          &   &  \\    
\hline 
\hline
\end{tabular}
\label{tab1}
\end{table}

\newpage
\begin{figure}[ht!]
\caption{Top view FESEM images of the GaN nanowall sample (a) before and (b) after 60~min of etching in aqueous solution of KOH. Respective inset shows the cross-sectional FESEM images. (c) Nanowall height $h$ as a function of etching time. (d) Average tip-width $t_{av}$ and the area fraction occupied by the tip of the walls $\phi$ as a function of etching time.}
\label{fig1}
\end{figure}

\begin{figure}[ht!]
\caption{(a) Four-probe resistivity $\rho_{FP}$ measured at 300~K  as a function of the height of the nanowalls. Inset shows the logarithm of  conductance $G$ = $\rho_{FP}$/$h$ as a function of $h$. (b) Conductivity at a height $h$ of the nanowall $\sigma(h)$ as a function of $t_{av}$. Inset schematically shows the contact pad arrangement on the sample surface (discussed in the text). }
\label{fig2}
\end{figure}   

\begin{figure}[ht!]
\caption{(a) Schematic design of the field effect transistor device used for the estimation of mobility through the nanowalls.(b) Source to drain current $I_{SD}$ as a function of gate to drain voltage $V_{GD}$ for the unetched nanowall sample. (c) $I_{SD}$ versus $V_{GD}$ characteristic for the sample after 110~min of etching.}
\label{fig3}
\end{figure} 

\begin{figure}[ht!]
\caption{(a) Change of conductance $\Delta G(B)$ = $G(B)-G(0)$ measured at 2~K as a function of the magnetic field $B$ for the unetched and the 60~min etched sample. (b) $\Delta G(B)$ as a function of $B$ in close-up for the 60~min etched sample. Solid lines represent the fit using Eq.~\ref{eq1}.}
\label{fig4}
\end{figure} 

\begin{figure}[ht!]
\caption{(a) Ga 3$d$ XPS spectra for the nanowall sample (solid black line) and a GaN epitaxial layer grown on c-plane sapphire substrate in the same growth chamber (red symbols). (b) schematically represents the band bending at one of the facets of the nanowall. (c) Theoretically calculated conduction band profile inside a wall of thickness $w$ = 20~nm and 40~nm. Inset compares the charge density profiles for $w$ = 20~nm and 40~nm.}
\label{fig5}
\end{figure}
\vfill

\newpage
\phantom{000}
\vfill
\centerline{\includegraphics*[width=8cm]{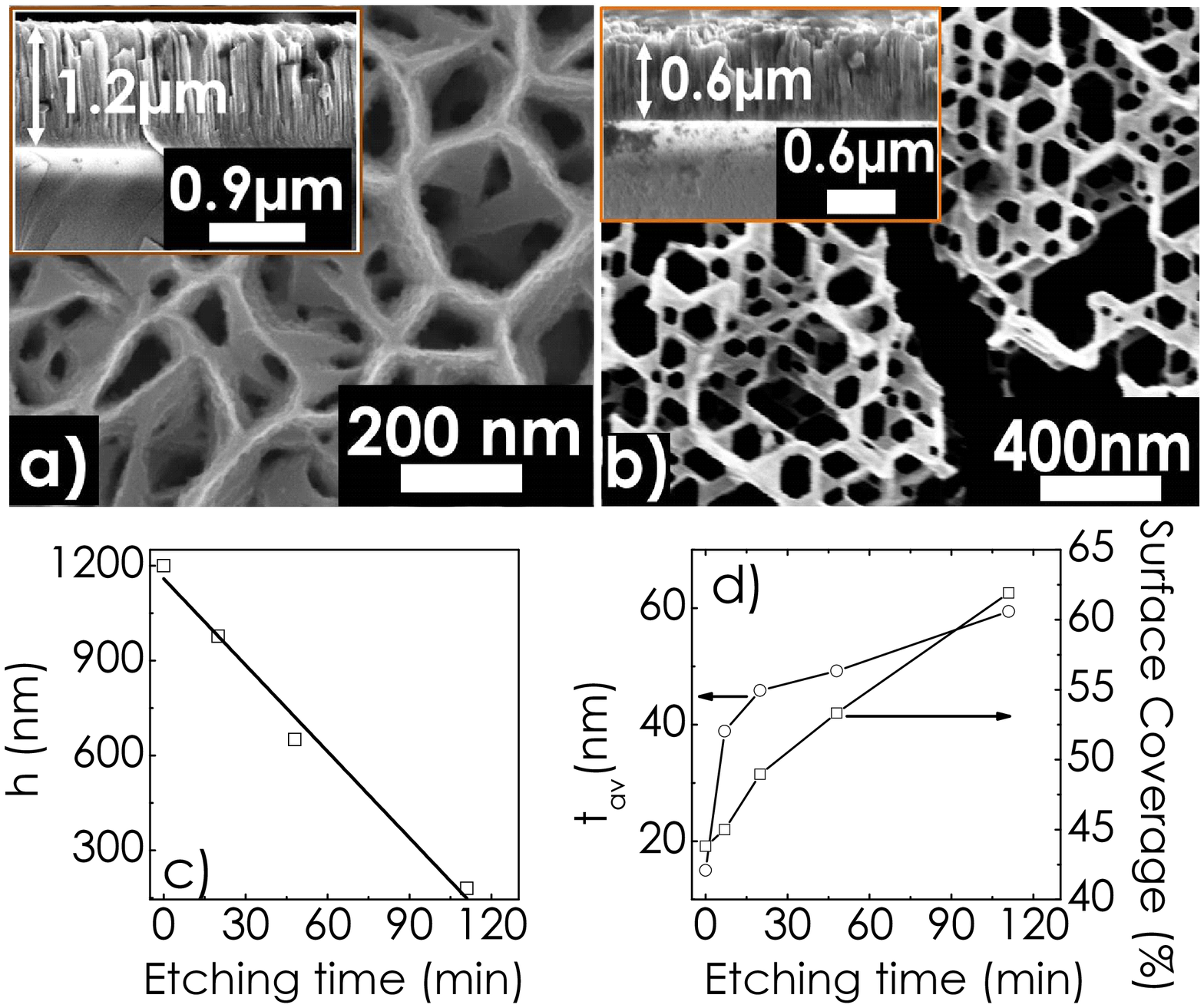}}
\vfill
\textsf{\large Fig.~1 of Bhasker \textsl{et al.}} 

\newpage
\phantom{000}
\vfill
\centerline{\includegraphics*[width=8cm]{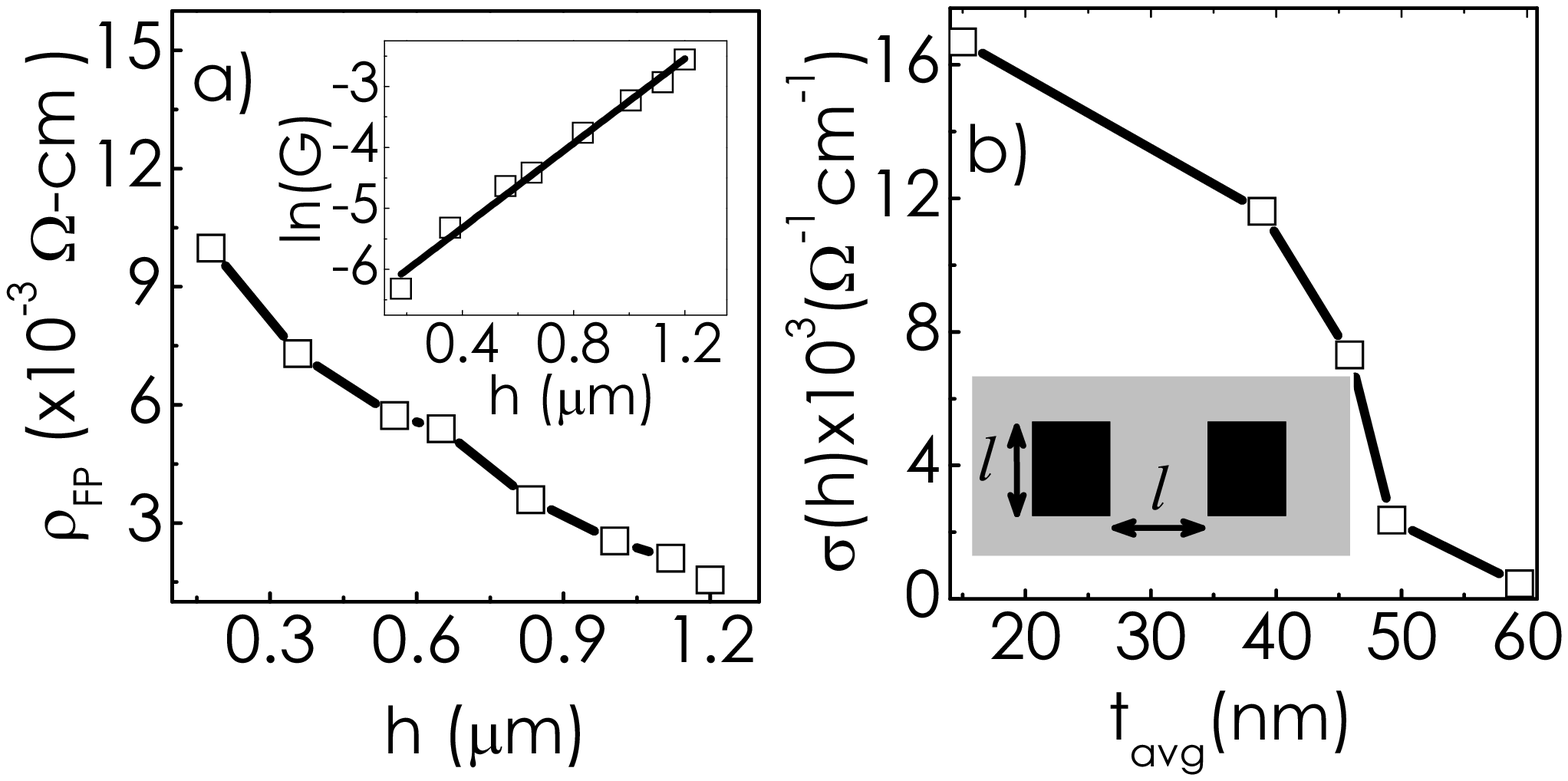}}
\vfill
\textsf{\large Fig.~2 of Bhasker \textsl{et al.}}

\newpage
\phantom{000}
\vfill
\centerline{\includegraphics*[width=8cm]{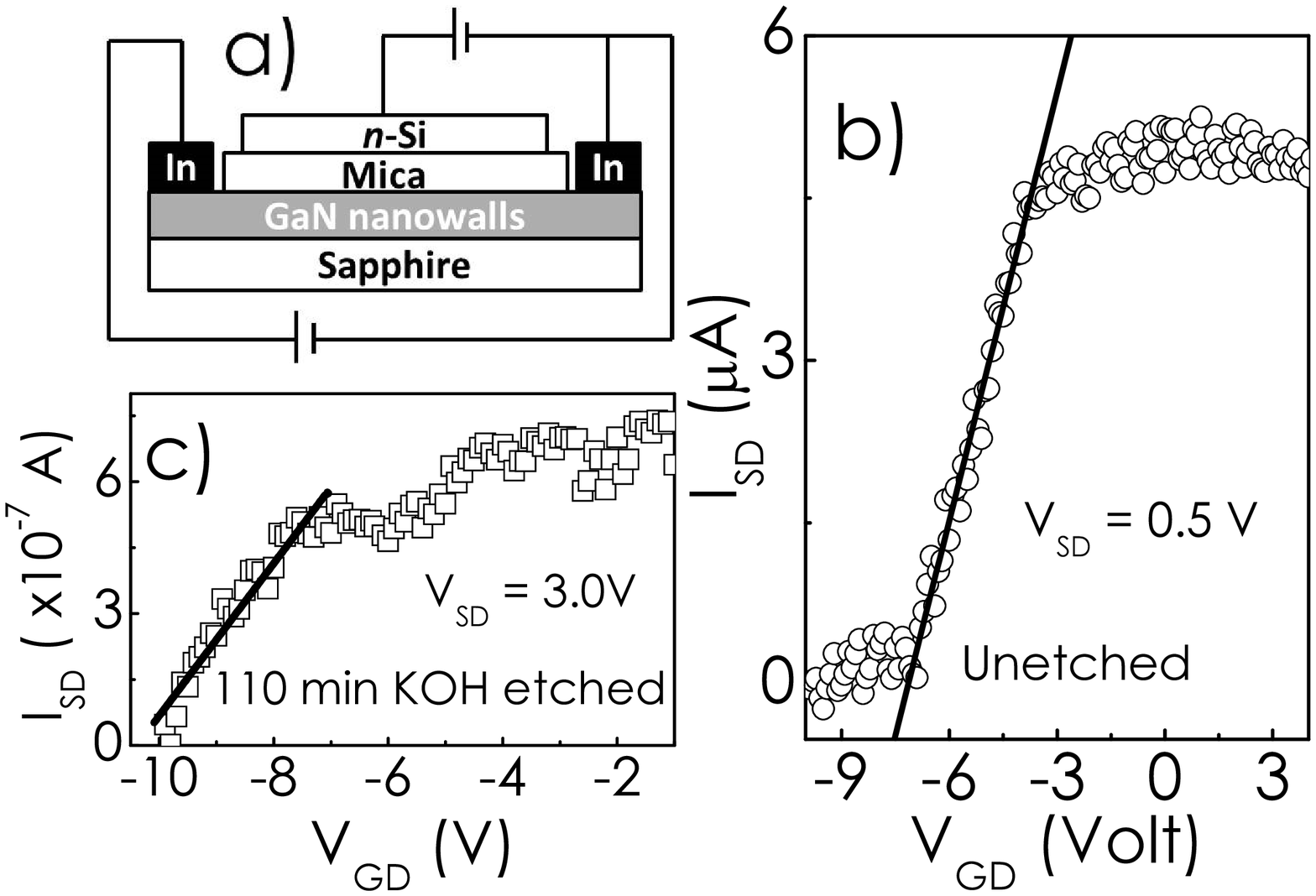}}
\vfill
\textsf{\large Fig.~3 of Bhasker \textsl{et al.}}

\newpage
\phantom{000}
\vfill
\centerline{\includegraphics*[width=8cm]{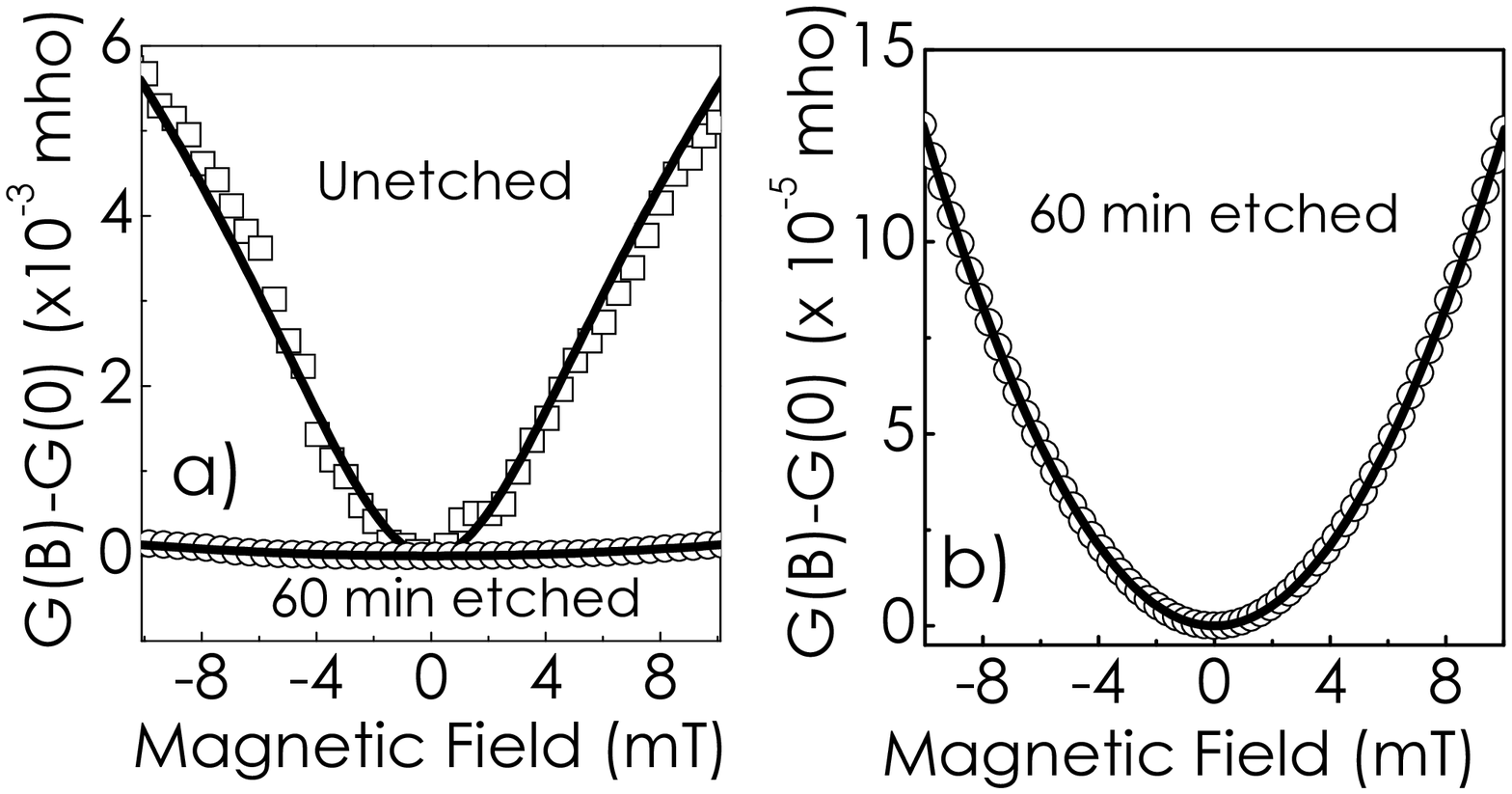}}
\vfill
\textsf{\large Fig.~4 of Bhasker \textsl{et al.}}

\newpage
\phantom{000}
\vfill
\centerline{\includegraphics*[width=8cm]{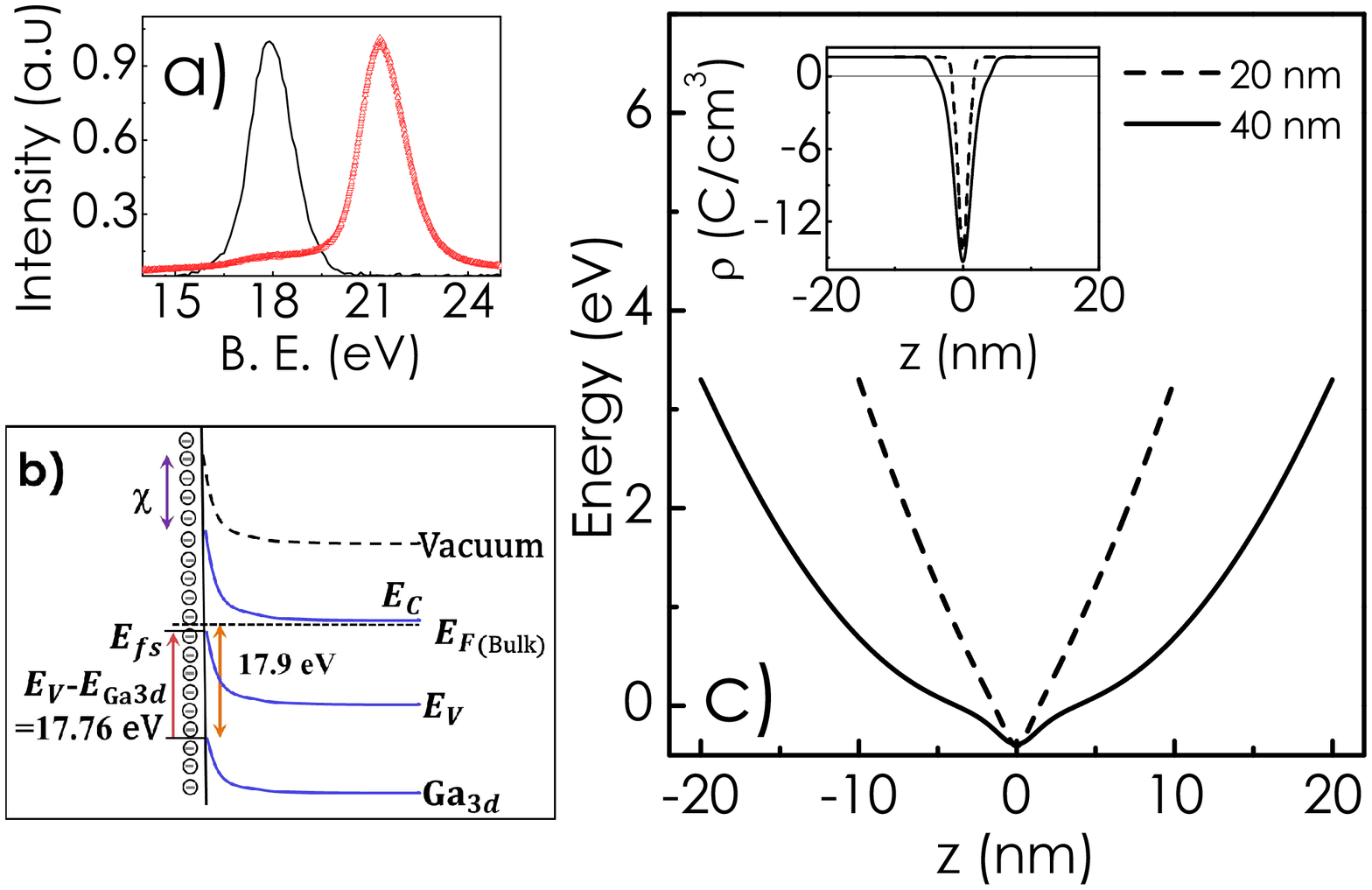}}
\vfill
\textsf{\large Fig.~5 of Bhasker \textsl{et al.}}


\begin{thebibliography}{99}
\bibitem{expl} The estimation of tip width is limited to the resolution of our FESEM. Cross-sectional TEM shows the tip width to be as small as 5~nm for some of the walls prior to the etching.
\bibitem{expl1} Much larger density of disorder is necessary to achieve weak localization in three dimension than in one or two dimension.
\end{thebibliography}
\end{document}